\documentclass[epj]{svjour}
\usepackage{graphics}
\usepackage[pdftex]{graphicx}
\usepackage{amsmath}
\usepackage{amssymb}
\begin{document}
\title{The $\eta^\prime$-carbon potential at low meson momenta}

\author{
M.~Nanova$^{1}$,~S.~Friedrich$^{1}$~V.~Metag$^{1}$,~E.~Ya.~Paryev$^{2}$,~F.~N.~Afzal$^{3}$,~D.~Bayadilov$^{3,4}$,~B.~Bantes$^{5}$,~R.~Beck$^{3}$,~M.~Becker$^{3}$,
S.~B\"ose$^{3}$,~K.-T.~Brinkmann$^{1}$,~V.~Crede$^{6}$,~P.~Drexler$^{1,a}$,~H.~Eberhardt$^{5}$,~D.~Elsner$^{5}$,~F.~Frommberger$^{5}$,~Ch.~Funke$^{3}$,\\~M.~Gottschall$^{3}$,~M.~Gr\"uner$^{3}$,~E.~Gutz$^{1}$,~Ch.~Hammann$^{3}$,~J.~Hannappel$^{5}$,~J.~Hartmann$^{3}$,~W.~Hillert$^{5,b}$,~Ph.~Hoffmeister$^{3}$,\\~Ch.~Honisch$^{3}$, T.~Jude$^{5}$,~D.~Kaiser$^{3}$,~F.~Kalischewski$^{3}$,~I.~Keshelashvili$^{7,c}$,~F.~Klein$^{5}$,~K.~Koop$^{3}$,~B.~Krusche$^{7}$,~M.~Lang$^{3}$,\\~K.~Makonyi$^{1,d}$,~F.~Messi$^{5}$,~J.~M\"uller$^{3}$,~J.~M\"ullers$^{3}$,~D.~Piontek$^{3}$,~T.~Rostomyan$^{7}$,~D.~Schaab$^{3}$,~Ch.~Schmidt$^{3}$,~H.~Schmieden$^{5}$, \\R.~Schmitz$^{3}$,~T.~Seifen$^{3}$,~V.~Sokhoyan$^{3,a}$,~C.~Sowa$^{8}$,~K.~Spieker$^{3}$,~A.~Thiel$^{3}$,~U.~Thoma$^{3}$,~T.~Triffterer$^{8}$,~M.~Urban$^{3}$, H.~van~Pee$^{3}$,~D.~Walther$^{3}$,~Ch.~Wendel$^{3}$,~D.~Werthm\"uller$^{7,e}$,~U.~Wiedner$^{8}$,~A.~Wilson$^{3}$,~L.~Witthauer$^{7}$,~Y.~Wunderlich$^{3}$, and~H.-G.~Zaunick$^{1}$\\
(The CBELSA/TAPS Collaboration)
\mail{Mariana.Nanova@exp2.physik.uni-giessen.de}}
\titlerunning{The $\eta^\prime$-carbon potential at low meson momenta}
\authorrunning{M. Nanova \textit{et al.}}

\institute{
{$^{1}$II. Physikalisches Institut, Universit\"at Gie{\ss}en, Germany}\\
{$^{2}$Institute of Nuclear Research, Russian Academy of Sciences, Moscow, Russia}\\
{$^{3}$Helmholtz-Institut f\"ur Strahlen- und Kernphysik, Universit\"at Bonn, Germany}\\
{$^{4}$Petersburg Nuclear Physics Institute, Gatchina, Russia}\\
{$^{5}$Physikalisches Institut, Universit\"at Bonn, Germany}\\
{$^{6}$Department of Physics, Florida State University, Tallahassee, FL, USA}\\
{$^{7}$Departement Physik, Universit\"at Basel, Switzerland}\\
{$^{8}$Physikalisches Institut, Universit\"at Bochum, Germany}\\
{$^{a}$Current address: Institut f\"ur Kernphysik, Universit\"at Mainz}\\
{$^{b}$Current address: Institute of Experimental Physics, University of Hamburg, Germany}\\
{$^{c}$Current address: Institut f\"ur Kernphysik, Forschungszentrum J\"ulich, Germany}\\
{$^{d}$Current address: Stockholm University, Stockholm, Sweden}\\
{$^{e}$Current address: School of Physics and Astronomy, University of Glasgow, UK}
}

\date{Received: date / Revised version: date}
%
\abstract{The production of $\eta^\prime$ mesons in coincidence with forward-going protons has been studied in photon-induced reactions on $^{12}$C and on a liquid hydrogen  (LH$_2$) target for incoming photon energies of 1.3-2.6 GeV at the electron accelerator ELSA. The $\eta^\prime$ mesons have been identified via the $\eta^\prime\rightarrow \pi^0 \pi^0\eta \rightarrow 6 \gamma$ decay registered with the CBELSA/TAPS detector system. Coincident protons have been identified in the MiniTAPS BaF$_2$ array at polar angles of $2^{\circ} \le \theta _{p} \le 11^{\circ}$. Under these kinematic constraints the $\eta^\prime$ mesons are produced with relatively low kinetic energy ($\approx$ 150 MeV) since the coincident protons take over most of the momentum of  the incident-photon beam. For the C-target this allows the determination of the real part of the $\eta^\prime$-carbon potential at low meson momenta by comparing with collision model calculations of the $\eta^\prime$ kinetic energy distribution and excitation function. Fitting the latter data for $\eta^\prime$ mesons going backwards in the center-of-mass system yields a potential depth of V = $-$(44 $\pm$ 16(stat)$\pm$15(syst)) MeV, consistent with earlier determinations of the potential depth in inclusive measurements for average $\eta^\prime$ momenta of $\approx$ 1.1 GeV/$c$. Within the experimental uncertainties, there is no indication of a momentum dependence of the $\eta^\prime$-carbon potential. The LH$_2$ data, taken as a reference to check the data analysis and the model calculations, provide differential and integral cross sections in good agreement with previous results for $\eta^\prime$ photoproduction off the free proton.}

%
\maketitle
\section{Introduction}
\label{intro}

The interaction of light pseudo-scalar mesons with nuclei has extensively been studied experimentally as well as theoretically as a test of Quantum  Chromodynamics in the strong coupling regime  \cite{HH,LMM,Oset,Zaki}. These studies are motivated by the possible existence of mesic states, i.e. meson-nucleus bound states. The existence of deeply-bound pionic states has been established  experimentally \cite{Gilg,Itahashi,Geissel,Suzuki}. In these systems, one electron in an inner orbit is replaced by a negatively-charged meson. These systems are bound by the attractive Coulomb interaction between a negatively-charged meson and the positively-charged nucleus. In case of the $\pi^-$, the superposition with the strong interaction, which is repulsive at low pion momenta, leads to a potential pocket near the nuclear surface and consequently to a halo-like $\pi^-$ distribution \cite{Kienle_Yamazaki}. In contrast, the nuclear interaction is attractive for $K^-$ mesons. However, the strongly-absorptive potential attenuates the atomic wave function within the interior of the nucleus and expels it to the nuclear surface, again leading to halo-like meson-nucleus configuration \cite{Friedman_Gal}.

For neutral mesons only the strong interaction can be an agent for the formation of meson-nucleus bound states. The question is whether the nuclear meson-nucleus interaction is attractive and sufficiently strong and whether the meson absorption in nuclei is sufficiently weak to allow for the formation of relatively narrow states. Following \cite{Nagahiro_PRC74}, the interaction of mesons with nuclei can be described by a potential
\begin{equation}
U(r) = V(r) + i W(r), 
\end{equation}
comprising a real part and an imaginary part, where $r$ is the distance of the meson to the centre of the nucleus. The depth of the real part $V(r)$ of the potential is a measure of the attraction and the size of the imaginary part $W(r)$ describes the strength of meson absorption. 

The $\eta^\prime$ - nucleon interaction is of scalar nature. A vector potential acts 
with a different sign for particles and antiparticles and is thus not allowed for particles that are their own antiparticles, like the $\eta^\prime$.
For a scalar meson-nucleus interaction, the depth of the real potential can then be related to the modification $\Delta m$ of the meson mass at normal nuclear matter density $\rho_0$  according to \cite{Metag_PPNP}
\begin{equation}
V(r) =  \Delta m \cdot c^2 \cdot \frac{\rho(r)}{\rho_0}.
\end{equation}
Here, $\rho(r)$ is the nuclear density profile, $\rho_0$ is the nuclear saturation density and $c$ the velocity of light.

The imaginary part of the potential describes the meson absorption in the medium via inelastic channels and is related to the in-medium width $\Gamma_{0}$ of the meson at nuclear saturation density by \cite{Nagahiro_PRC74}
\begin{equation}
W(r) = -\frac{1}{2}\Gamma_{0}\cdot \frac{\rho(r)}{\rho_{0}}.
\end{equation}

There are two conditions for the existence and experimental observation of meson-nucleus bound states: (i) the real part of the potential should be sufficiently deep, in particular for very small meson momenta near the production threshold; (ii) a small imaginary potential implies a narrow width allowing for an easier separation of signal and background and thus an easier identification of the bound state. Furthermore, the width should be smaller than the spacing of bound states to avoid overlapping levels. The imaginary part of the potential should therefore be small compared to the real potential, i.e. $\vert W \vert \ll \vert V \vert $. 

The strength of the real and imaginary parts of the $\eta^\prime$-nucleus potential has been determined by the CBELSA/ TAPS collaboration in a series of inclusive $\eta^\prime$ photoproduction experiments at the electron accelerator ELSA \cite{Nanova_C,Nanova_Nb,Nanova_TA,Metag_Hypint,Nanova_Metag,Friedrich_TA}. The depth of the real potential has been extracted from measurements of the excitation function, i.e. the cross section for $\eta^\prime$ production as a function of the incident photon energy, and from measurements of the momentum differential $\eta^\prime$ production cross section. The potential parameters are deduced from a comparison of measured with corresponding calculated cross sections, obtained with collision model calculations \cite{Paryev} for different potential depths. These model calculations describe the production of mesons in proton-, photon- and pion-induced reactions off nuclei, using the elementary production cross sections as input. These investigations consider direct and two-step production mechanisms and take the internal nucleon momentum distributions - including high momentum tails - into account. The off-shell propagation of the produced mesons is approximated by assuming a density-averaged modified in-medium meson mass. For average $\eta^\prime$ momenta comparable to the $\eta^\prime$ mass, the depth of the real potential has been found to be $V_0 = V(\rho= \rho_0) = -(39\pm7$(stat)$\pm15$(syst)) MeV \cite{Nanova_Nb}. 

The imaginary potential has been extracted from measurements of the transparency ratio, defined in \cite{Hernandez_Oset,Cabrera_NPA733} as
\begin{equation}
T_A = \frac{\sigma_{\gamma A \rightarrow m X}}{A \cdot \sigma_{\gamma N \rightarrow m X}}\label{eq:transp}.
\end{equation}
It compares the production cross section per nucleon of meson $m$ off a nucleus with mass number A with the production cross section on a free nucleon $N$.  $T_A$ quantifies the loss of the meson flux in a nuclear target through inelastic reactions which are related to the imaginary part of the meson in-medium self-energy or width. The imaginary potential is deduced from a comparison of measured with calculated transparency ratios obtained in transport- \cite{Buss,Weil} or collision-model calculations \cite{Paryev}, assuming different in-medium width or absorption cross sections for the meson. The imaginary part of the $\eta^\prime$ - nucleus potential has been found to be in the range of $\approx$ -(7.5 -12.5) MeV \cite{Nanova_C}. The transparency ratio has not only been measured as a function of the nuclear mass number $A$ but also as a function of the $\eta^\prime$ momentum. Extrapolating to near threshold momenta, an imaginary potential of $W_0 = W(\rho = \rho_0) = -(13\pm3$(stat)$\pm3$(syst)) MeV has been deduced in \cite{Friedrich_TA}. The modulus of the imaginary potential is thus about a factor 3 smaller than the modulus of the real potential, indicating favourable conditions for the observability of $\eta^\prime$-nucleus bound states, provided this ratio of imaginary to real potential persists also at low momenta which are decisive for the possible formation of $\eta^\prime$ - nucleus bound states. The real part of the $\eta^\prime$ -nucleus potential has, however, not yet been determined for meson momenta small compared to the meson mass, while the imaginary part of the potential is known for small momenta. The motivation for the present work is to provide equivalent information on the real part of the $\eta^\prime$ - nucleus potential at near-threshold momenta. 

Low-momentum $\eta^\prime$ mesons are selected in the experiment by requiring the participant proton to be emitted at forward angles. The forward-going proton takes over most of the momentum of the incoming photon beam; the $\eta^\prime$ meson then goes backward in the $\gamma$-proton center-of-mass system but, boosted into the laboratory system, the $\eta^\prime$ meson moves slowly forward in the laboratory, providing the possibility to study the $\eta^\prime$ meson-nucleus interaction at very low momenta $p_{\eta^\prime}$ compared to the $\eta^\prime$ meson mass $m_{\eta^\prime}$: ($p_{\eta^\prime} \ll m_{\eta^\prime}$). The real part of the $\eta^\prime$-nucleus potential is then extracted by comparing the measured kinetic energy distribution and excitation function for $\eta^\prime$ mesons in coincidence with forward-going protons with corresponding calculations within a collision model \cite{Paryev_peta}. A corresponding measurement on the $\omega$ meson has been reported in \cite{Friedrich_PLB}.

It should be noted that for short-lived mesons like the $\eta^\prime$ meson, external meson beams cannot be produced and experiments as described here are the only way to obtain information on the meson-nucleus interaction. The nucleus serves as a production target but is simultaneously used to probe the meson-nucleus interaction. Hereby it is important that the nucleus does not disintegrate in the meson production process. This can be ensured by appropriate experimental conditions and event selection criteria, as described below.

The paper is structured as follows: The experimental setup and the conditions of the experiment are described in section 2. Details of the analysis are given in section 3. Section 4 presents the experimental results and the comparison with the mentioned theoretical calculations. Concluding remarks are given in section 5.

\section{Experimental setup}
\label{expsetup}
The experiment was performed at the electron stretcher accelerator ELSA in Bonn \cite{Husmann_Schwille,Hillert}. Photons were produced by scattering electrons of 3.2~GeV off a 50-$\mu$m-thick copper radiator and impinged on a 5-mm-thick carbon target, corresponding to 5.9$\%$ of a radiation length $X_0$. For the photoproduction off the free proton a 500-$\mu$m-thick diamond radiator and a 5-cm-long LH$_2$ target were used. The bremsstrahlung photons were tagged in the energy range of 0.7-3.1~GeV by detecting the scattered electrons in coincidence after deflection by a tagging magnet. Decay photons from $\eta^\prime$ mesons produced by the interaction in the target were detected with the combined Crystal Barrel (CB) (1320 CsI(Tl) modules) \cite{Aker} and MiniTAPS calorimeters (216 BaF$_2$ modules) \cite{TAPS1,TAPS2}. This detector setup covered polar angles of 11$^{\circ}$-156$^{\circ}$ and 1$^{\circ}$-11$^{\circ}$, respectively, and the full azimuthal angular range, thereby covering 96$\%$ of the full solid angle. In the angular range of 11$^\circ$-28$^\circ$ the CB modules 
were read out by photomultipliers, providing energy and time information while the rest of the CB crystals were read out by photodiodes with energy information only. Because of the high granularity and the large solid-angle coverage the detector system was ideally suited for the detection and reconstruction of multi-photon events. 

At polar angles of $1^{\circ}$-$11^{\circ}$, protons were registered in plastic scintillators in front of the MiniTAPS forward wall and identified by time-of-flight measurements and their energy depositions in the BaF$_2$ modules of MiniTAPS. In the angular range of 11$^{\circ}$-28$^{\circ}$ charged particles were registered in plastic scintillators in front of the CB modules and for 23$^{\circ}$-156$^{\circ}$ they were identified in a three-layer scintillating fibre array \cite{Suft}. The polar angular resolution for proton detection is $\sigma =1^{\circ}$ in MiniTAPS and $\sigma =6^{\circ}$ in 11$^{\circ}$-156$^{\circ}$, given by the size of the CsI crystals.\\

The photon flux through the target was determined by counting the photons reaching the Gamma Intensity Monitor (GIM) \cite{GIM} at the end of the setup in coincidence with electrons registered in the tagging system. The total rates in the tagging system were $\approx$10~MHz for the C experiment and $\approx$17~MHz for the measurement with the free proton target. The polarization of the incident-photon beam in the LH$_2$ experiment, obtained by using the diamond crystal as a bremsstrahlung target, was not exploited in the analysis of these data. During the carbon experiment an aerogel-Cherenkov detector with a refractive index of n=1.05 was used to veto electrons, positrons and charged pions in the angular range covered by MiniTAPS. This device was replaced for the LH$_2$ beamtime with a gas-Cherenkov detector with a refractive index of n=1.00043 in order to veto electrons and positrons. The data were collected during two data-taking periods of 525~h for the carbon and 330~h for the LH$_2$ target. 

 \begin{figure*}
\begin{center}
 \resizebox{0.8\textwidth}{!}
 {\includegraphics[width=5.5cm,clip]{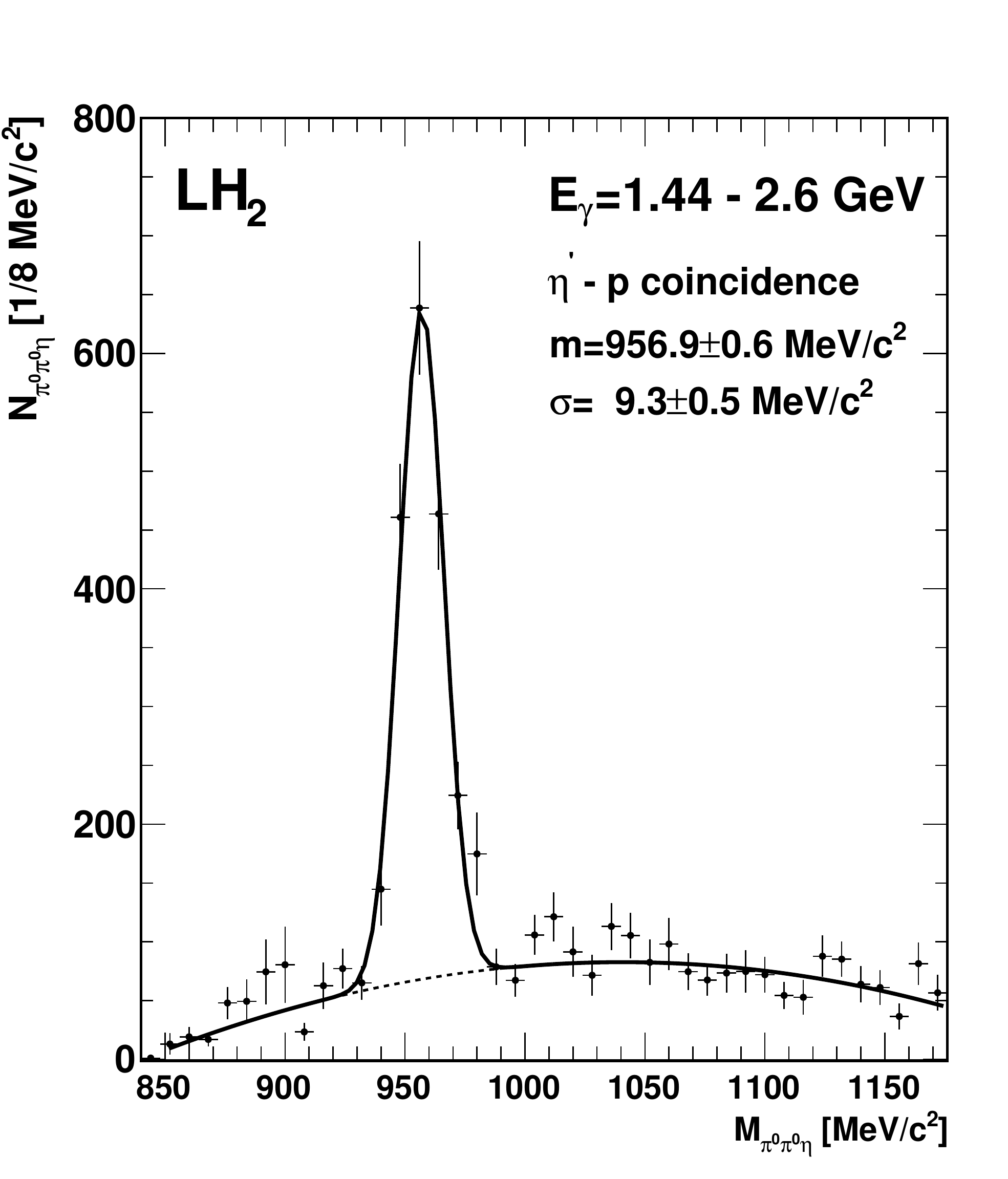} \includegraphics[width=5.5cm,clip]{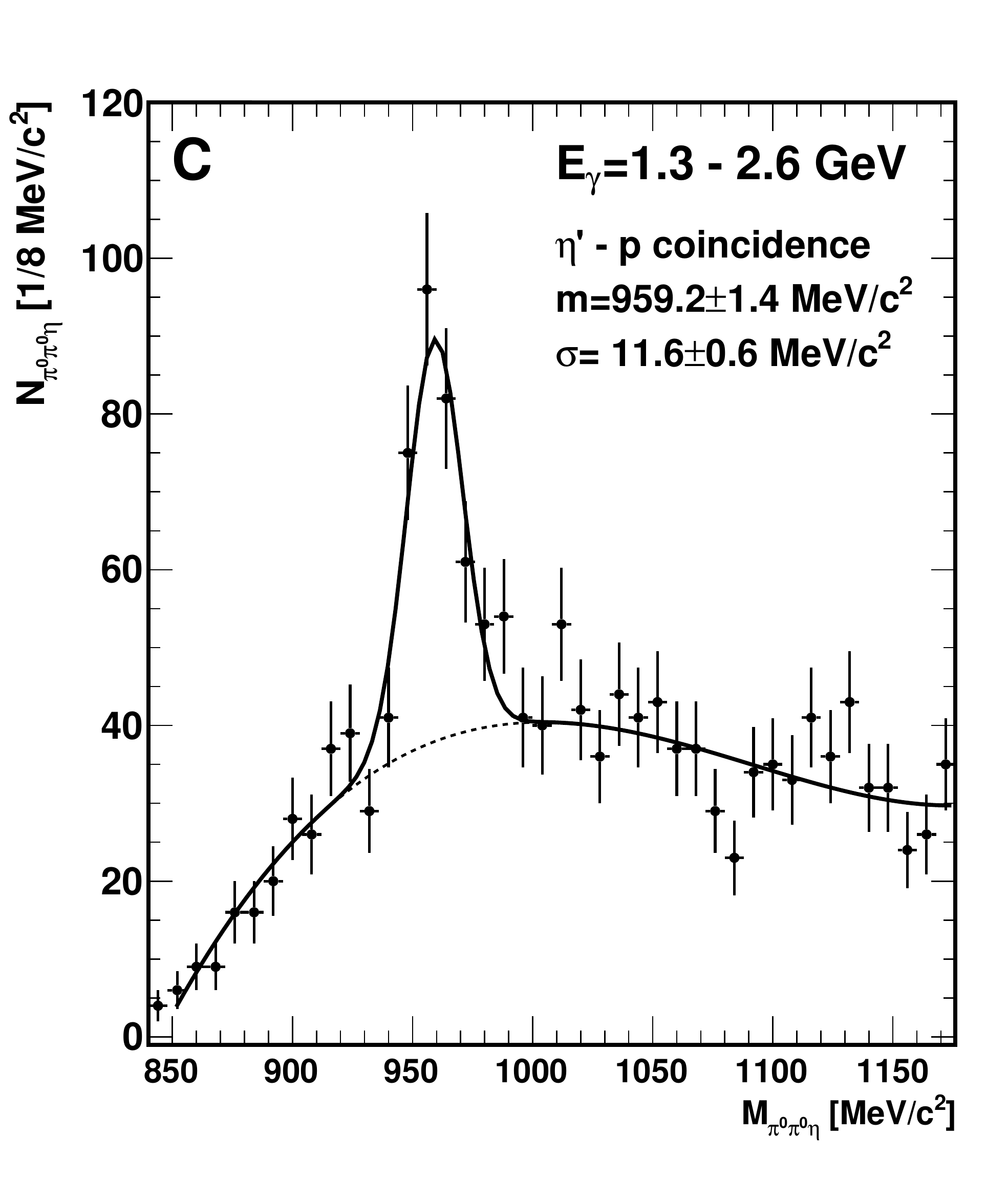}}
 \caption{$\pi^0 \pi^0 \eta$ invariant mass distribution in coincidence with a proton registered in MiniTAPS for the LH$_2$ (left) and carbon (right) targets.The fitted mass peak positions are consistent with the $\eta^\prime$ mass quoted by the particle data group \cite{PDG}. The relative mass resolutions are 1.0$\%$ and 1.2 $\%$ for the LH$_2$ and carbon target, respectively.}
\label{fig:invmass}
\end{center}
\end{figure*}
\begin{figure*}
\begin{center}
 \resizebox{1.0\textwidth}{!}
 { \includegraphics[width=8cm,clip]{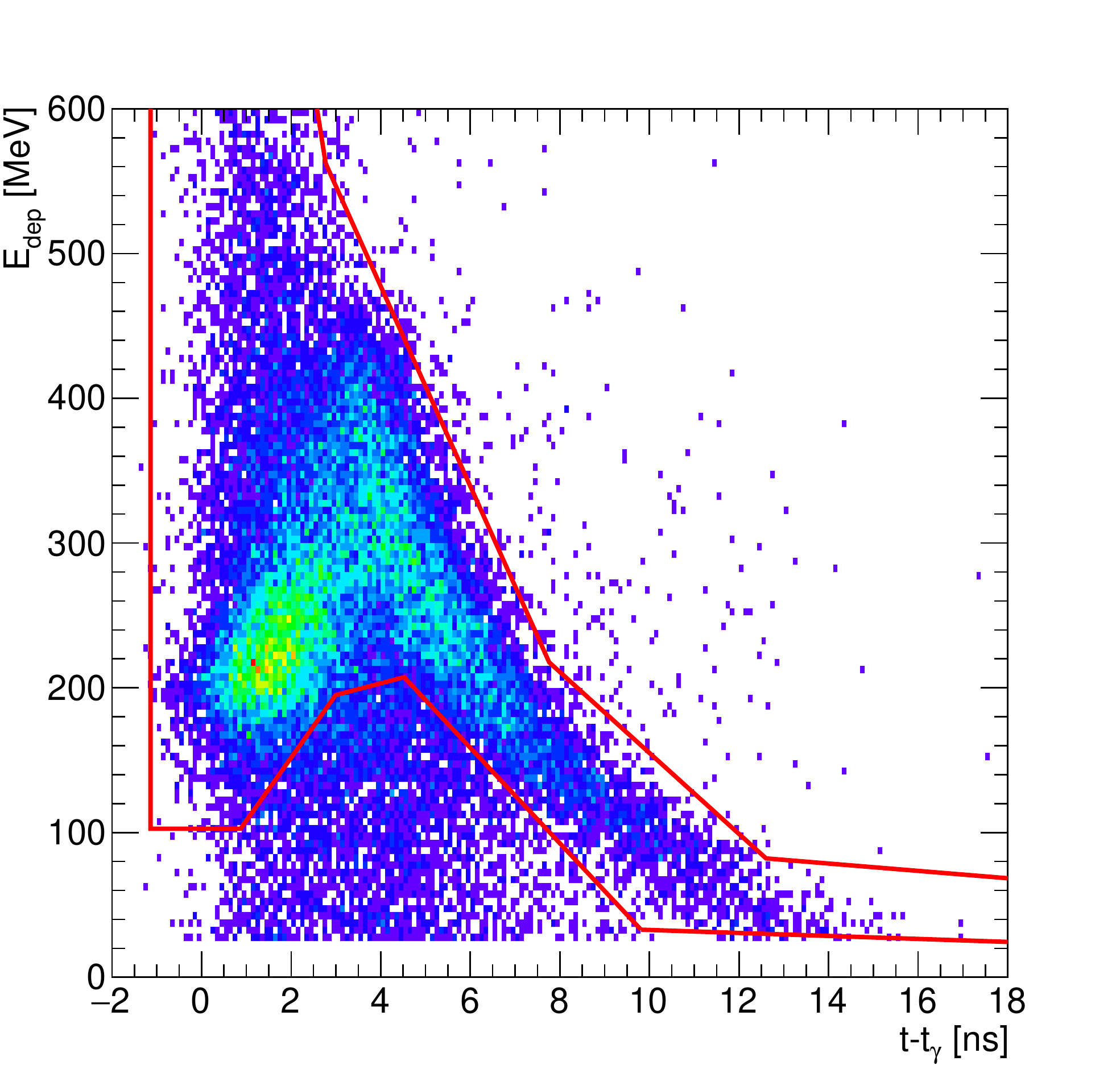} \includegraphics[width=8cm,clip]{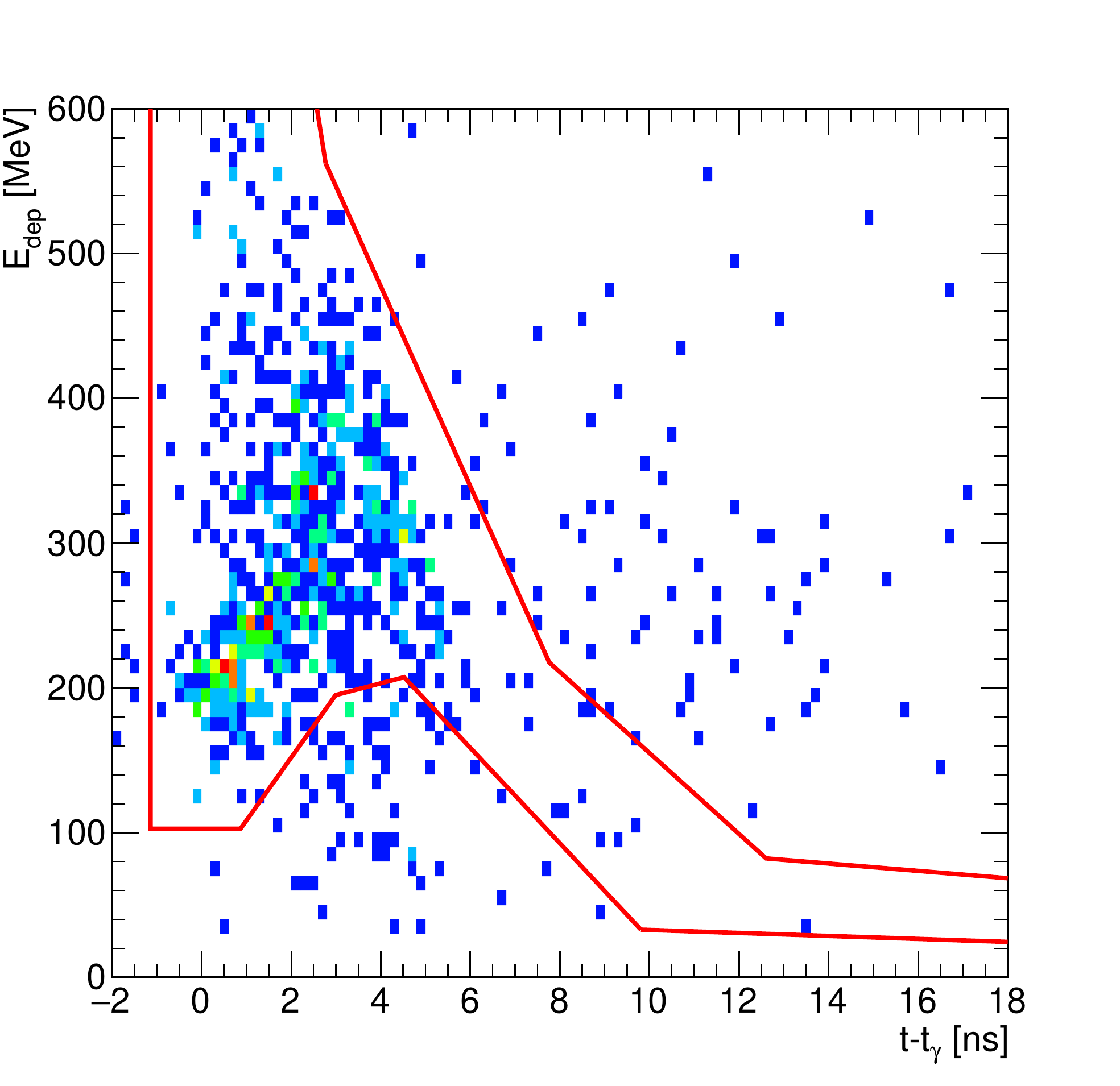}  \includegraphics[width=8cm,clip]{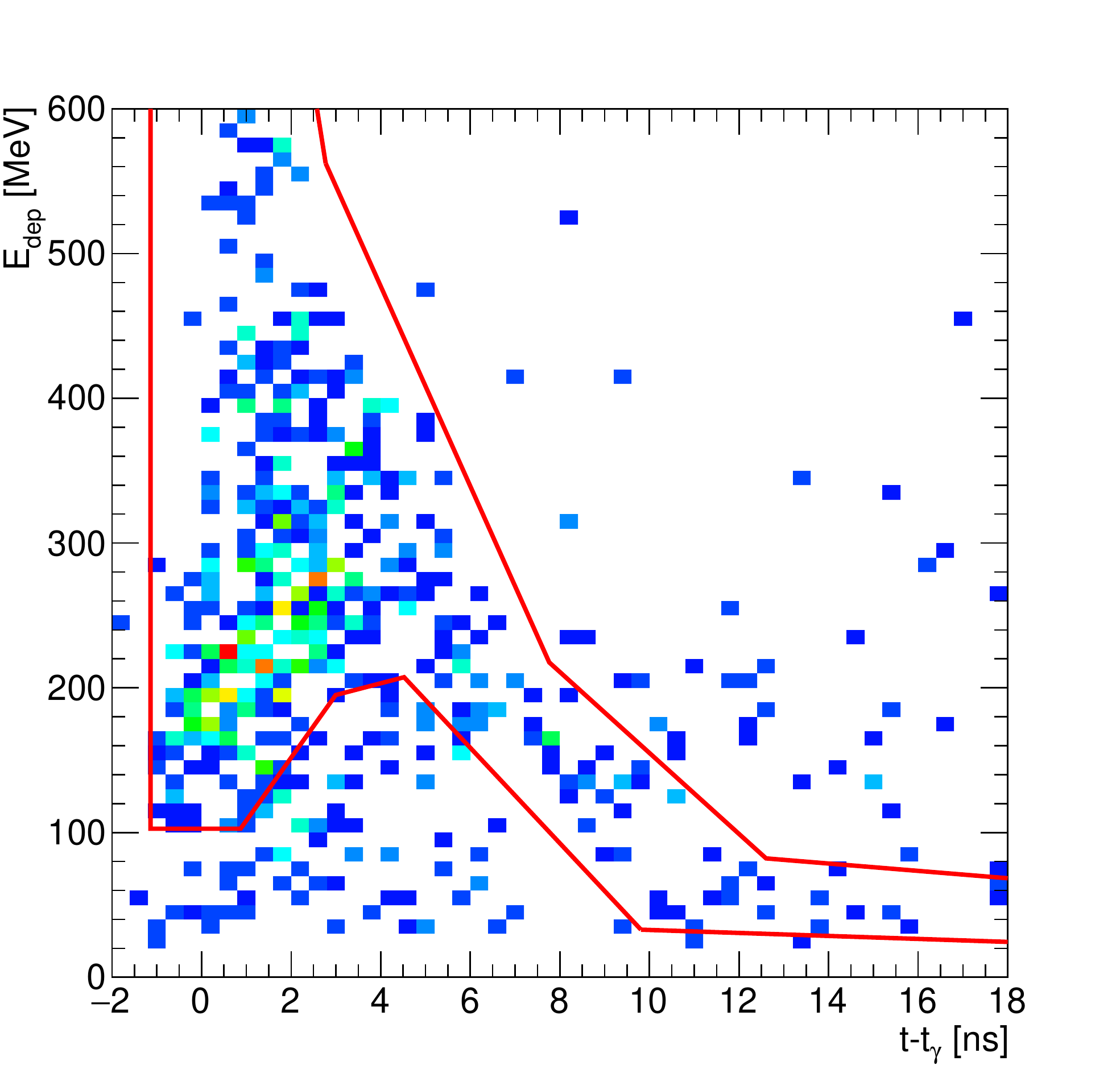}}
 \caption{Time-of-flight of charged particles relative to photons versus energy deposited in the MiniTAPS BaF$_2$ detectors: (left) GEANT3 simulation for protons from the $\gamma C \rightarrow p \eta^\prime X$ reaction; data from the LH$_2$ (center) and carbon experiment (right), requesting the coincident $\eta^\prime$ meson to go backward in the center-of-mass system. The two dimensional cut for identifying protons is indicated.} 
\label{fig:pid}
\end{center}
\end{figure*} 

The $\eta^\prime$ mesons were identified via the $\eta^\prime\rightarrow \pi^0\pi^0\eta\rightarrow 6\gamma$ decay chains, which have a total branching ratio of 8.5$\%$ \cite{PDG}. In the carbon experiment, the first-level trigger selected events with at least four hits in the combined electromagnetic calorimeters, requiring in addition that the aerogel-Cherenkov detector had not fired (veto-condition); in the LH$_2$ experiment, a less restrictive trigger was applied, requiring two or more hits in the calorimeters and no hit in the gas-Cherenkov detector. The dead time introduced by the Cherenkov detectors was about 10$\%$ for the aerogel-Cherenkov detector and 4$\%$ for the gas-Cherenkov detector. The photon flux has been corrected for the GIM dead time which was about 13$\%$ in the carbon experiment and 25$\%$ in the LH$_2$ experiment. A more detailed description of the detector setup and the running conditions can be found in \cite{Friedrich_PLB,GIM,Afzal}. 

 \begin{figure*}
\begin{center}
 \resizebox{0.9\textwidth}{!}
 { \includegraphics[width=8.5cm,clip]{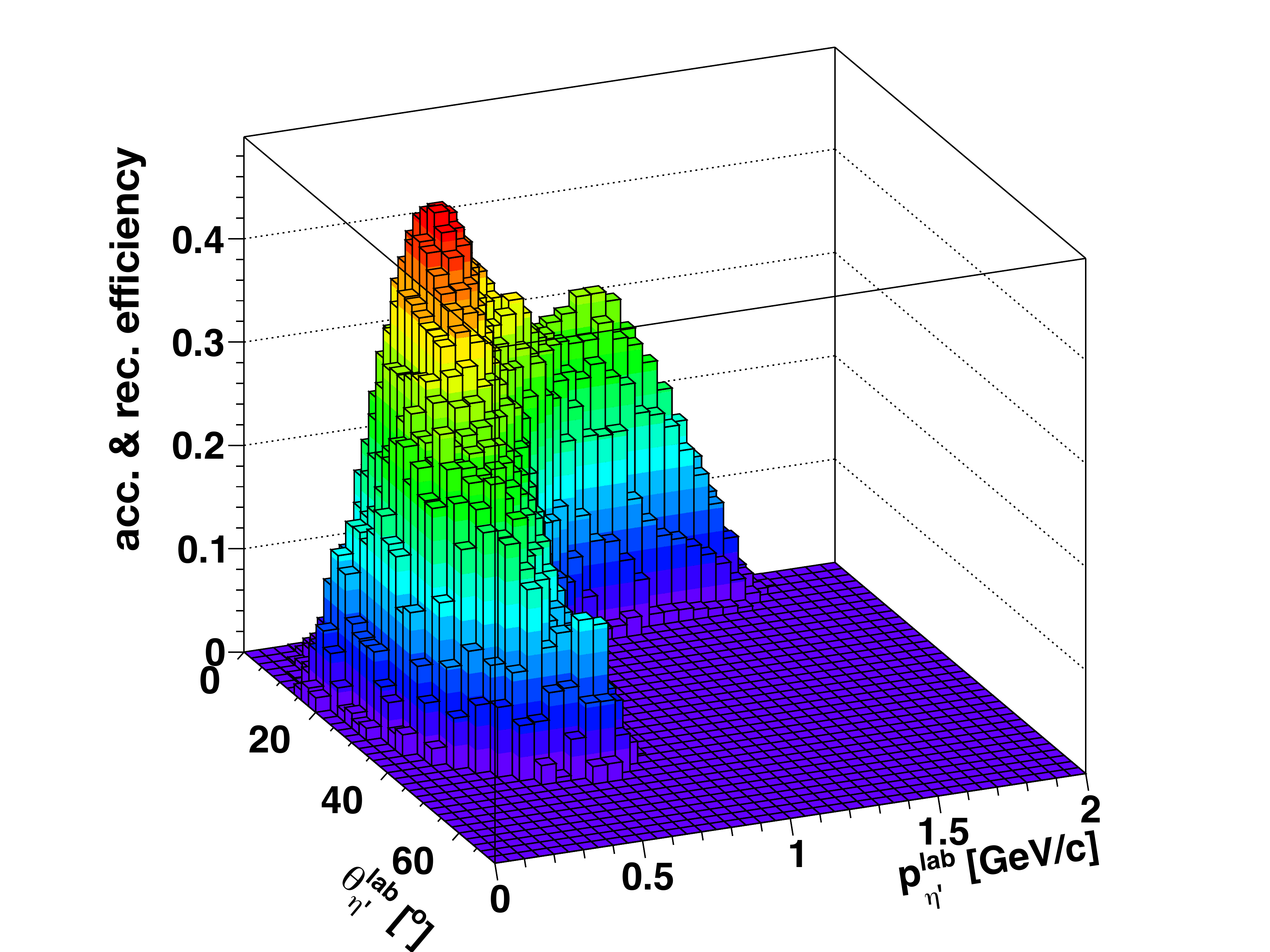}\includegraphics[width=8.5cm,clip]{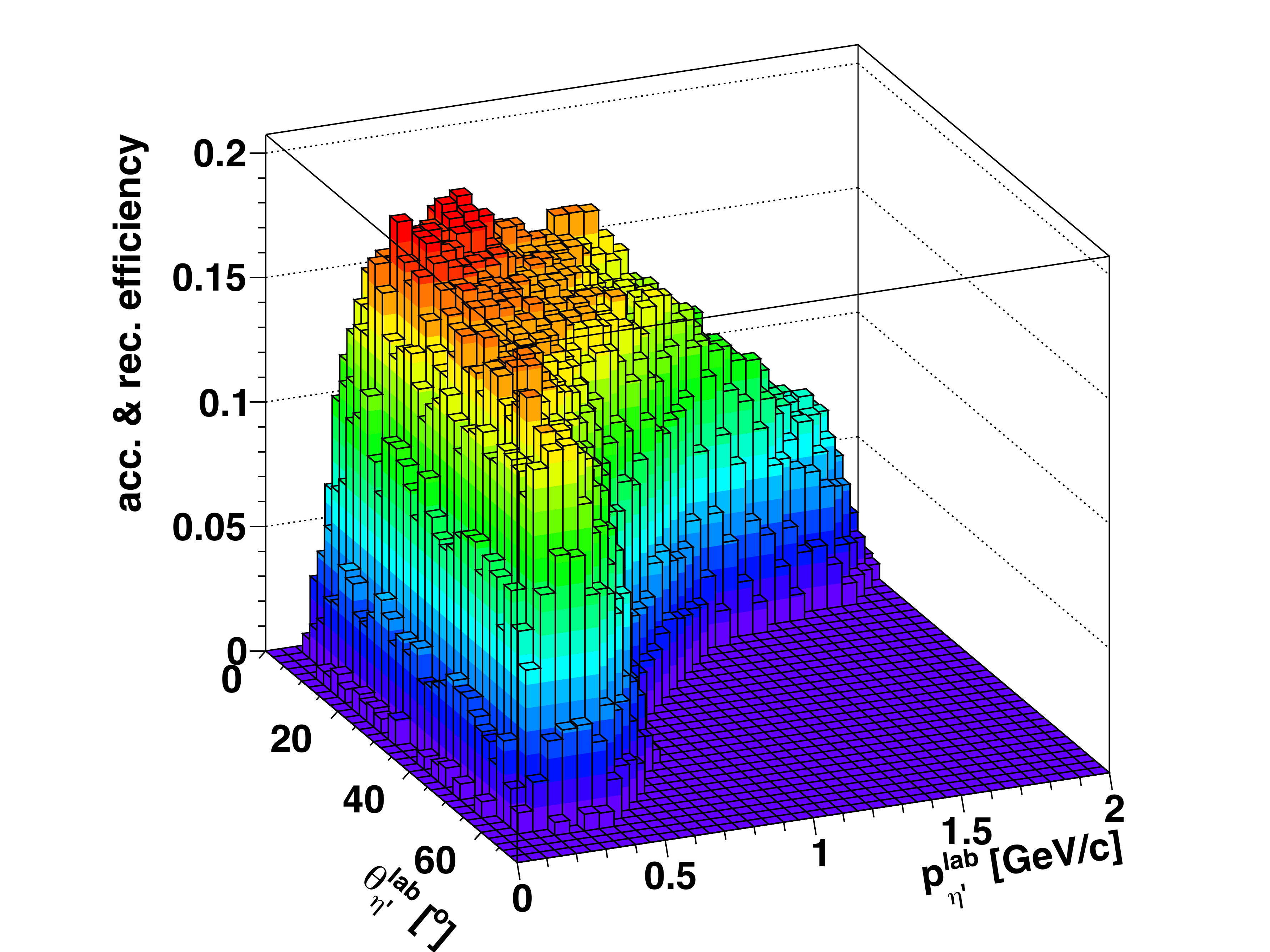}}
 \caption{Acceptance and reconstruction efficiency of $\eta^\prime$ - proton coincidences for photoproduction off the free proton target (left) and the carbon target (right) in the incident photon energy range of 1.44 - 2.6 GeV (LH$_{2}$) and 1.3 - 2.6~GeV (C), respectively. The  $\eta^\prime$ mesons are registered in the $\eta^\prime \rightarrow \pi^0 \pi^0 \eta \rightarrow 6 \gamma$ decay channel and protons are identified by time-of-flight and the energy deposited in the MiniTAPS BaF$_2$ detectors (see Fig.~\ref{fig:pid}). The acceptance and reconstruction efficiency is shown for all $\eta^\prime$ momenta although only those for slow $\eta^\prime$ mesons with $p_{\eta^\prime} \lesssim m_{\eta^\prime}$ are effectively used.} 
\label{fig:acc}
\end{center}
\end{figure*}

 \section{Data analysis}
\label{sec:ana}
In the off-line analysis, events of interest were selected and the background was suppressed by several kinematical cuts. Only events with incident photon energies in the range of  
1.3 - 2.6~GeV for the C target and 1.44 - 2.6 GeV for the LH$_{2}$ target were processed. Photons were required to have energies larger than 25~MeV to suppress cluster split-offs. Random coincidences between the tagger and the detector modules in the first-level trigger were removed by a cut in the corresponding time spectra around the prompt peaks and by sideband subtraction.

Events with one and only one charged hit and exactly 6 photons with an energy sum larger than 600~MeV were selected. The one charged hit had to be in MiniTAPS. Since the detector system covered almost the full solid angle, this condition suppressed all events more violent than quasi-free $\eta^\prime$ production processes which are characterised by a higher multiplicity of hits in the detector. Because of the proton detected in MiniTAPS, the residual nucleus most likely was $^{11}$B. Following a very simplified argument, it should be noted that in the initial stage of the reaction both, the $\eta^\prime$ meson and the proton are still in the C-nucleus; thus the $\eta^\prime$ mesons is subject to the $\eta^\prime$ - C potential. Since for the chosen kinematics, the proton is almost twice as fast as the $\eta^\prime$ meson (see Fig.~\ref{fig:pid} (right) and Fig.~\ref{fig:Ekin_Exfunc_C} (left)) the proton leaves the nucleus before the $\eta^\prime$ meson, so that then the $\eta^\prime$ meson probes the $\eta^\prime$ - B potential. Consequently, the measurement is sensitive to a mixture of both potentials. Because of the weak nuclear mass dependence of the potential parameters observed in \cite{Metag_PPNP,Nanova_C,Nanova_Nb} it is assumed in the data analysis - see below - that within the experimental errors there is no difference between the $\eta^\prime$-B and $\eta^\prime$-C potential.

The 6 photons were combined in two photon pairs with invariant masses in the range 110~MeV/$c^2 \le m_{\gamma\gamma} \le$ 160~MeV/$c^2$ (corresponding to a $\pm$3$\sigma$ cut around $m_{\pi^{0}}$) and one pair with invariant mass in the range \\500~MeV/$c^2 \le m_{\gamma\gamma} \le$ 600~MeV/$c^2$ (roughly corresponding to a $\pm$2$\sigma$ cut around $m_{\eta}$). The best photon combination was selected based on a $\chi^2$ minimization. To suppress the background from $\eta \rightarrow 3\pi^{0}$ decays and direct 3$\pi^{0}$ production, events with 3 $\gamma$ pairs, each one with an invariant mass within the limits for the pion mass ($m_{\pi^{0}}$) given above, were removed from the data set. The resulting $\pi^0\pi^0\eta$ invariant mass spectra obtained for both targets are shown in Fig.~\ref{fig:invmass}. 

In the data analysis, protons were requested in the angular range of $2^{\circ}-11^{\circ}$ covered by MiniTAPS where protons can be identified by requiring a signal in a plastic detector in front of a MiniTAPS-BaF$_2$ module as well as by time-of-flight and an energy deposition in this BaF$_2$ module consistent with the simulated detector response.  Fig.~\ref{fig:pid} shows the deposited energy versus the time-of-flight of charged hits in coincidence with $\pi^0 \pi^0 \eta$ events with an invariant mass of 930-990 MeV for both targets, corresponding to a $\pm 3\sigma$ cut around the $\eta^\prime$ mass. The proton identification cut was based on GEANT3 simulations shown in Fig.~\ref{fig:pid} (left). Protons were fully stopped in the BaF$_2$ modules up to a kinetic energy of about 400 MeV. For higher kinetic energies only a fraction of the energy was deposited decreasing with energy according to the Bethe-Bloch formula down to the energy deposition of $\approx$ 180 MeV for minimum-ionizing particles. Furthermore, choosing protons in the angular range of $2^{\circ}-11^{\circ}$  guaranteed coincident low-momentum $\eta^\prime$ mesons with  $p_{\eta^\prime} \lesssim m_{\eta^\prime}$. 

The acceptance and reconstruction efficiency for $\eta^\prime$ mesons in coincidence with protons in the polar angle range of $2^{\circ}-11^{\circ}$ was determined by Monte Carlo simulations. In the event generator, the measured angular differential cross sections for $\eta^\prime$ mesons produced off the free proton \cite{Crede} and off the proton or neutron bound in the deuteron \cite{Jaegle} were used as input, respectively. For the analysis of the carbon data the Fermi motion of nucleons in the target nucleus was taken into account by the parametrization proposed in \cite{Ciofi}. Photons from  $\eta^\prime$ decays and the recoil proton emerging from the centre of the target were tracked with the GEANT3 package \cite{GEANT} based on a full implementation of the detector system, including the response of BaF$_2$ detectors to protons. The combined effect of acceptance and reconstruction efficiency was determined as a function of the $\eta^\prime$  momentum and angle in the laboratory frame by taking the ratio of the number of reconstructed to the number of generated meson - proton coincidence events for each $\eta^\prime$ momentum and angular bin, including all selection cuts mentioned above. The resulting two-dimensional acceptance and reconstruction efficiency distributions for $\eta^\prime$ - proton coincidences are shown in Fig.~\ref{fig:acc} for both targets and the full meson-momentum ranges. The distributions vary smoothly as a function of the $\eta^\prime$ laboratory angle and momentum. For the analysis of the C-data the 2-dimensional reconstruction efficiency determination has the advantage that distortions of the meson angle and momentum due to final state interactions in the nucleus are directly taken into account by using the reconstruction efficiency for the observed final state meson momentum and angle. In the Monte Carlo simulations the same trigger conditions as in the experiment were applied, requiring $\ge$ 4 hits in the whole detector system for the C beamtime and $\ge$ 2 hits for the LH$_2$ experiment.

For both targets the same analysis procedure was applied. The $\eta^\prime$ kinetic energy distribution and the excitation function were derived by fitting the $\eta^\prime$ yields in $\pi^0\pi^0\eta$ mass spectra generated for 9 (7) kinetic energy bins and 7 (18) bins in the incident photon energy for the C (LH$_2$) analyses. The bin sizes were chosen according to the available statistics. When incrementing the $\pi^0\pi^0\eta$ invariant mass histograms, each event was weighted with the inverse photon flux at the given incident photon energy and the acceptance and reconstruction efficiency for an $\eta^\prime$-proton coincidence at the observed meson momentum and angle in the laboratory frame (see Fig.~\ref{fig:acc}). The $\eta^\prime$ yields were extracted by fitting the invariant mass spectra with a Gaussian line shape function together with a polynomial to describe the background distribution. The statistical errors, including the error of the random timing background subtraction, were obtained with standard techniques of error propagation in weighted histograms.

The different sources of systematic errors for the cross section determination are summarised in Table \ref{tab:syst}. The systematic errors in the fit procedure were estimated to be of the order of 10$\%$ by applying different background functions and fit intervals. The systematic errors of the the acceptance and reconstruction efficiency were determined to be less than 10$\%$ by varying the start distributions in the  simulation between isotropic and forward peaking $\eta^\prime$ angular distributions, as observed experimentally with increasing incident beam energy. Systematic errors associated with the photon flux determination using the GIM were estimated to be about 5-10$\%$. The systematic errors introduced by uncertainties in the absorption of incident photons in the C-target (photon shadowing) \cite{Falter,bianchi,Muccifora} (see below) were $\approx 5\%$. Adding the systematic errors quadratically, the total systematic error of the cross section determinations was $\approx$ 17$\%$. 
\begin{table}[h!]
\centering
\caption{Sources of systematic errors for cross section determination.}
\begin{footnotesize}
\begin{tabular}{cc}
fits & $\approx$ 10$\%$\\
reconstruction efficiency & $\lesssim$10$\%$\\
photon flux & 5-10$\%$\\
photon shadowing & $\approx$ 5$\%$\\
\hline
total  & $\approx$ 17$\%$\\
\end{tabular}
\end{footnotesize}
\label{tab:syst}
\end{table}

\section{Experimental results}
\begin{figure*}
\begin{center}
 \resizebox{0.9\textwidth}{!}
 { \includegraphics[width=6.5cm,clip]{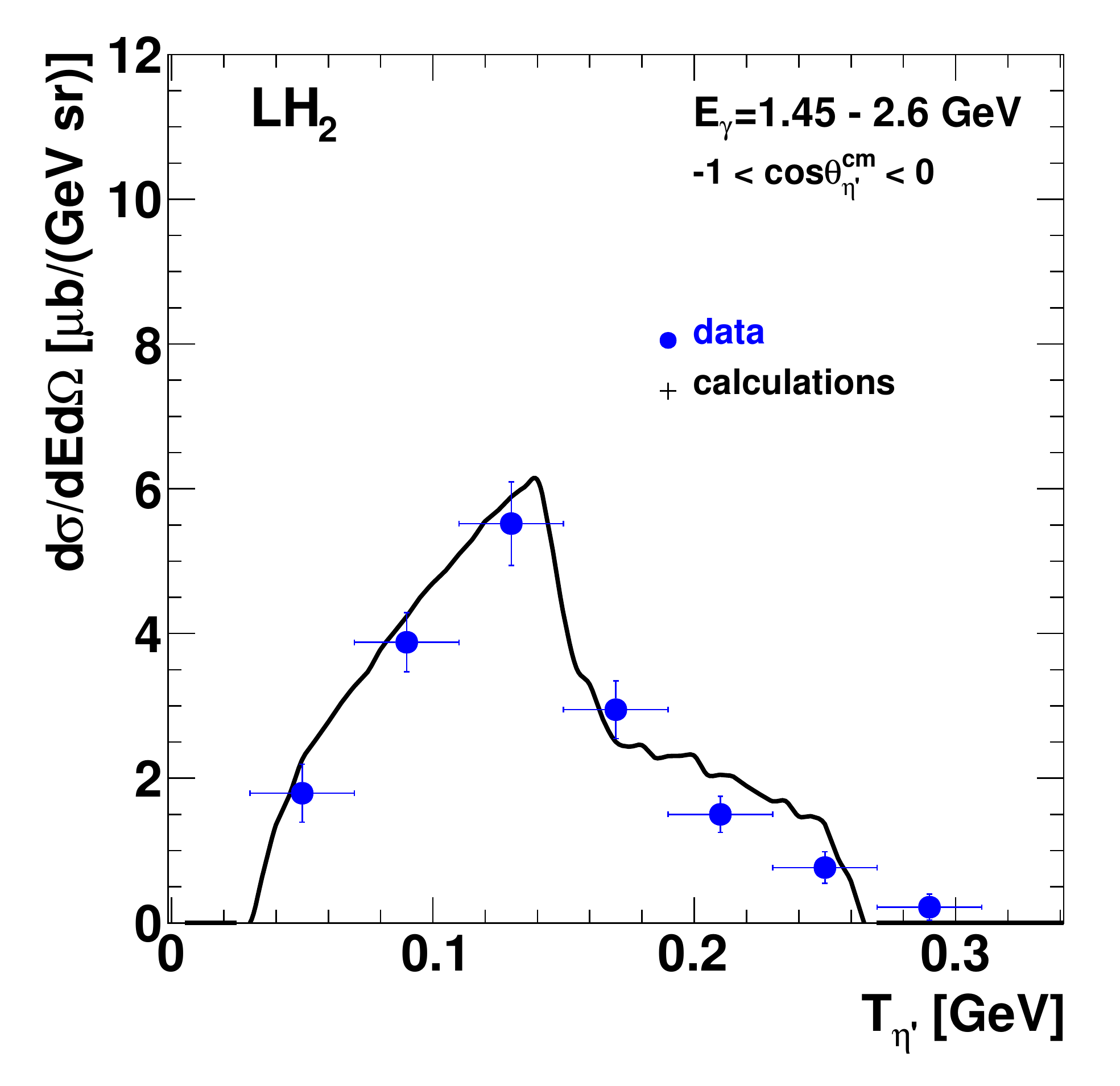}  \includegraphics[width=6.5cm,clip]{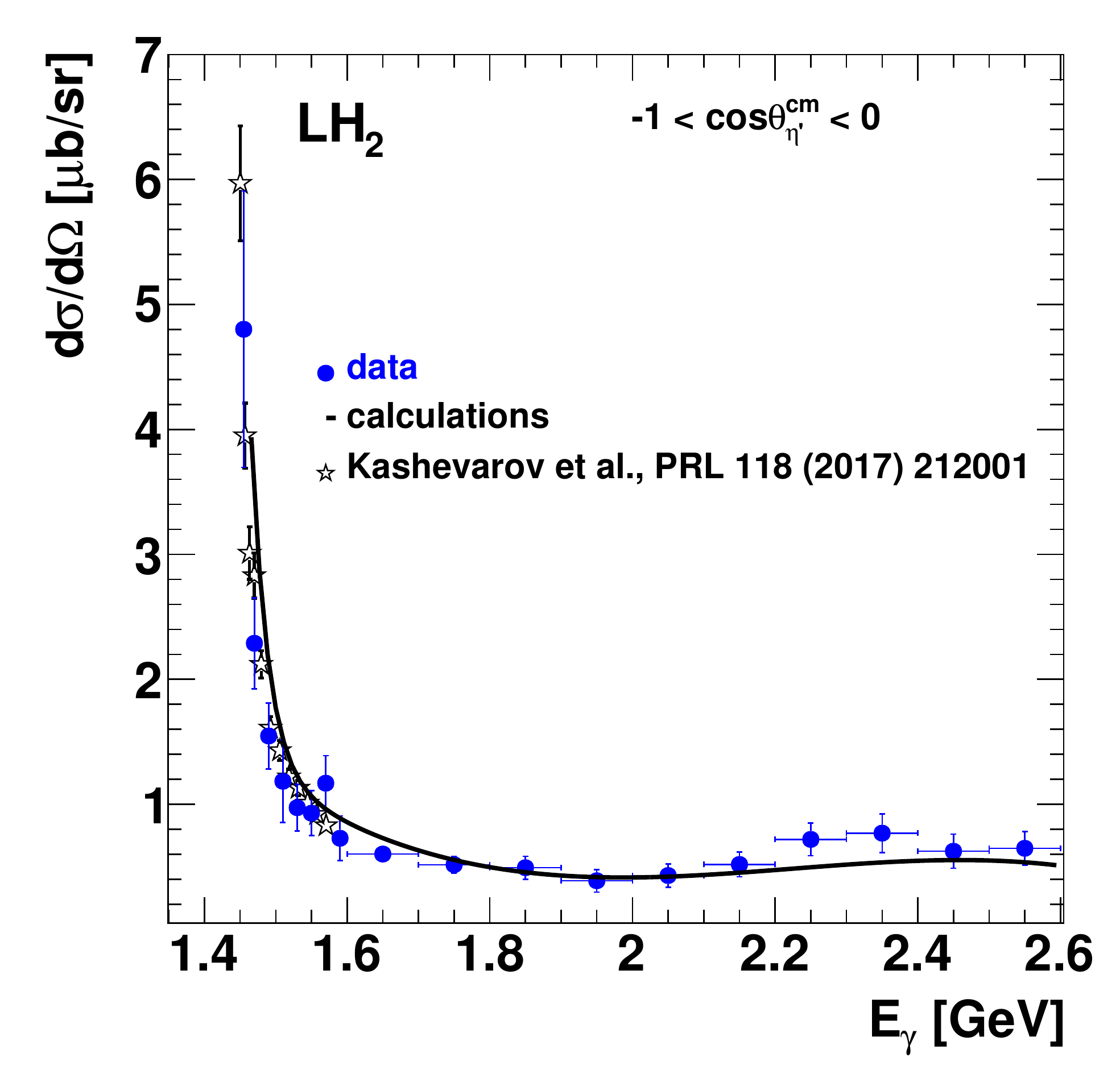}}
 \caption{Kinetic energy differential cross section integrated over photon energies 1.45-2.6 GeV (left) and cross section as a function of the incident photon energy (right) for $\eta^\prime$ mesons produced off the free proton and going backwards in the center-of-mass system. The $\eta^\prime$ mesons are detected in coincidence with protons registered in the polar angular range of $2^{\circ}-11^{\circ}$. The curves represent collision model calculations, using as input measured differential cross sections for $\eta^\prime$ photoproduction off the free proton \cite{Crede}; stars correspond to recent inclusive data for $\eta^\prime$ photoproduction off the free proton \cite{Kashevarov} converted into the laboratory system under the given kinematic constraints.}
 \label{fig:Ekin_Exfunc_LH2}
\end{center}
\end{figure*}

Figure~\ref{fig:Ekin_Exfunc_LH2} presents the energy differential cross section integrated over the photon energy range of 1.45-2.6 GeV (left) and the cross section as a function of the incident photon energy (right) for $\eta^\prime$ photoproduction off the free proton in coincidence with protons in the polar angle range of  $2^{\circ}-11^{\circ}$. In addition it is required that the $\eta^\prime$ mesons go backwards in the center-of-mass system. 

The data are compared to collision model calculations under the same kinematic conditions as in the experiment. These calculations are based on measured angular differential cross sections for $\eta^\prime$ mesons produced off the free proton \cite{Crede}. In addition, the data are compared to recent high statistics inclusive data for near threshold photoproduction of $\eta^\prime$ mesons \cite{Kashevarov} which have been converted into cross sections in the laboratory system for the given kinematics. Our data are in good agreement with the calculations and the recent experimental results demonstrating the reliability of the analysis procedure and model calculations.

\begin{figure*}
\begin{center}
 \resizebox{0.9\textwidth}{!}
 { \includegraphics[width=6.5cm,clip]{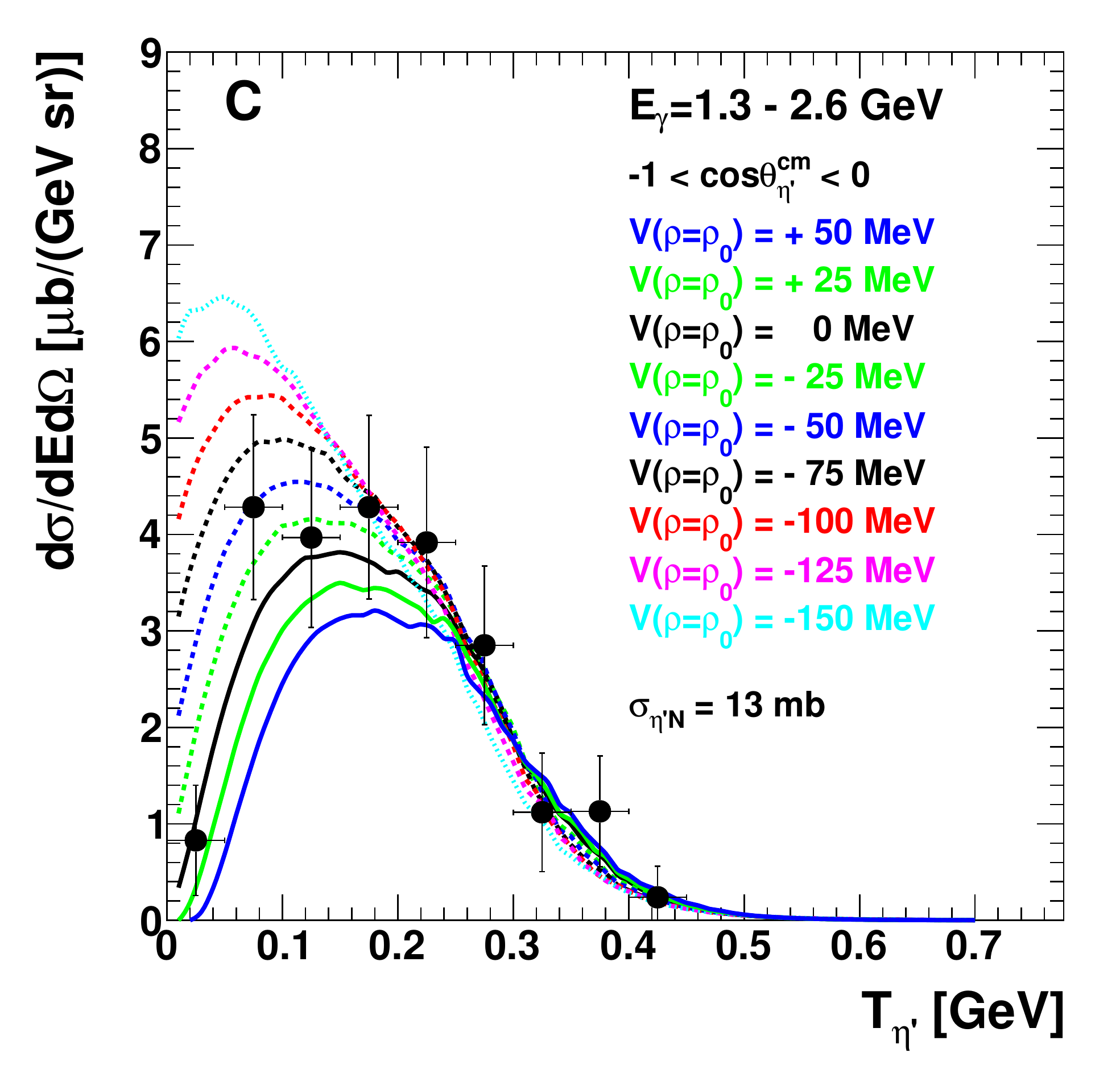}  \includegraphics[width=6.5cm,clip]{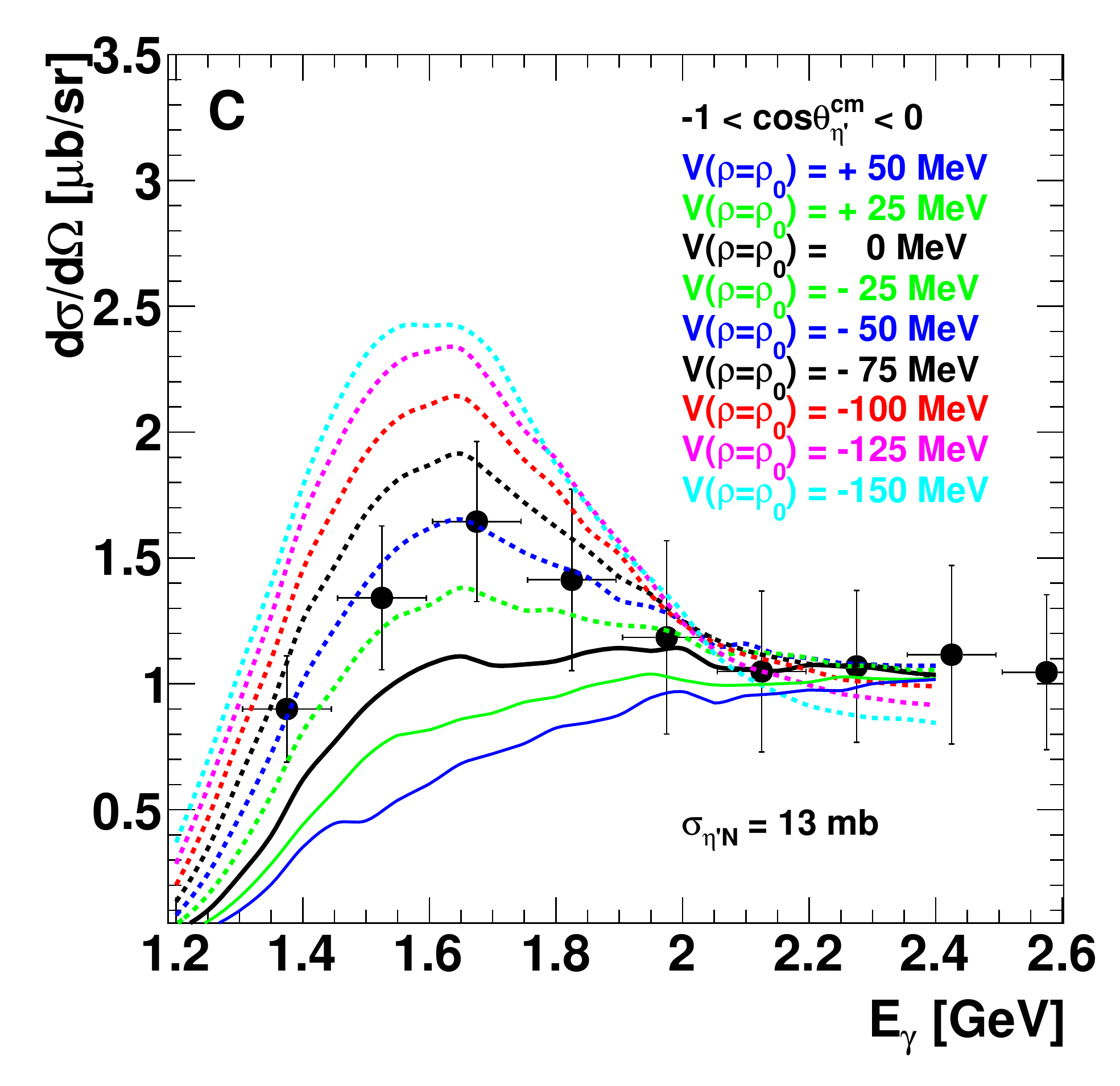}}
 \caption{Kinetic energy differential cross section integrated over photon energies 1.3-2.6 GeV (left) and excitation function (right) for $\eta^\prime$ mesons produced off the carbon target and going backwards in the center-of-mass system of the incident photon beam and a nucleon at rest. The $\eta^\prime$ mesons are detected in coincidence with protons registered in the polar angular range of $2^{\circ}-11^{\circ}$. The curves represent collision model calculations \cite{Paryev_peta} for the kinematic conditions of the experiment for different $\eta^\prime$  - carbon potentials, ranging from +50 to $-$150 MeV. The calculations have been performed for an in-medium inelastic $\eta^\prime-N$ cross section  of 13 mb, as determined in \cite{Friedrich_TA}.}
 \label{fig:Ekin_Exfunc_C}
\end{center}
\end{figure*}

Figure~\ref{fig:Ekin_Exfunc_C} presents the corresponding results for the carbon target obtained under the same kinematic conditions. 
 The $\eta^\prime$ kinetic energy distribution exhibits a broad maximum at 50-200 MeV and the excitation function shows a maximum around 1.7 GeV incident photon energy.  The average $\eta^\prime$ momentum is 600 MeV/$c$, i.e. about a factor 2 smaller than the average $\eta^\prime$ momentum in the inclusive experiments \cite{Nanova_C,Nanova_Nb}. For the carbon target the lowest data point in the kinetic energy distribution corresponds to an $\eta^\prime$ momentum of 220 MeV/$c$, comparable to the Fermi momentum of nucleons in carbon. The excitation function for $\eta^\prime$ -proton coincidences, i.e. the differential cross section as a function of the incident photon energy, is shown in Figure~\ref{fig:Ekin_Exfunc_C} (right). The cross sections include a 15$\%$ correction for absorption of the incoming photon beam (photon sha\-dow\-ing) for carbon \cite{Falter,bianchi,Muccifora}.
\begin{figure*}
\begin{center}
 \resizebox{1.0\textwidth}{!}
  { \includegraphics[width=11.0cm,clip]{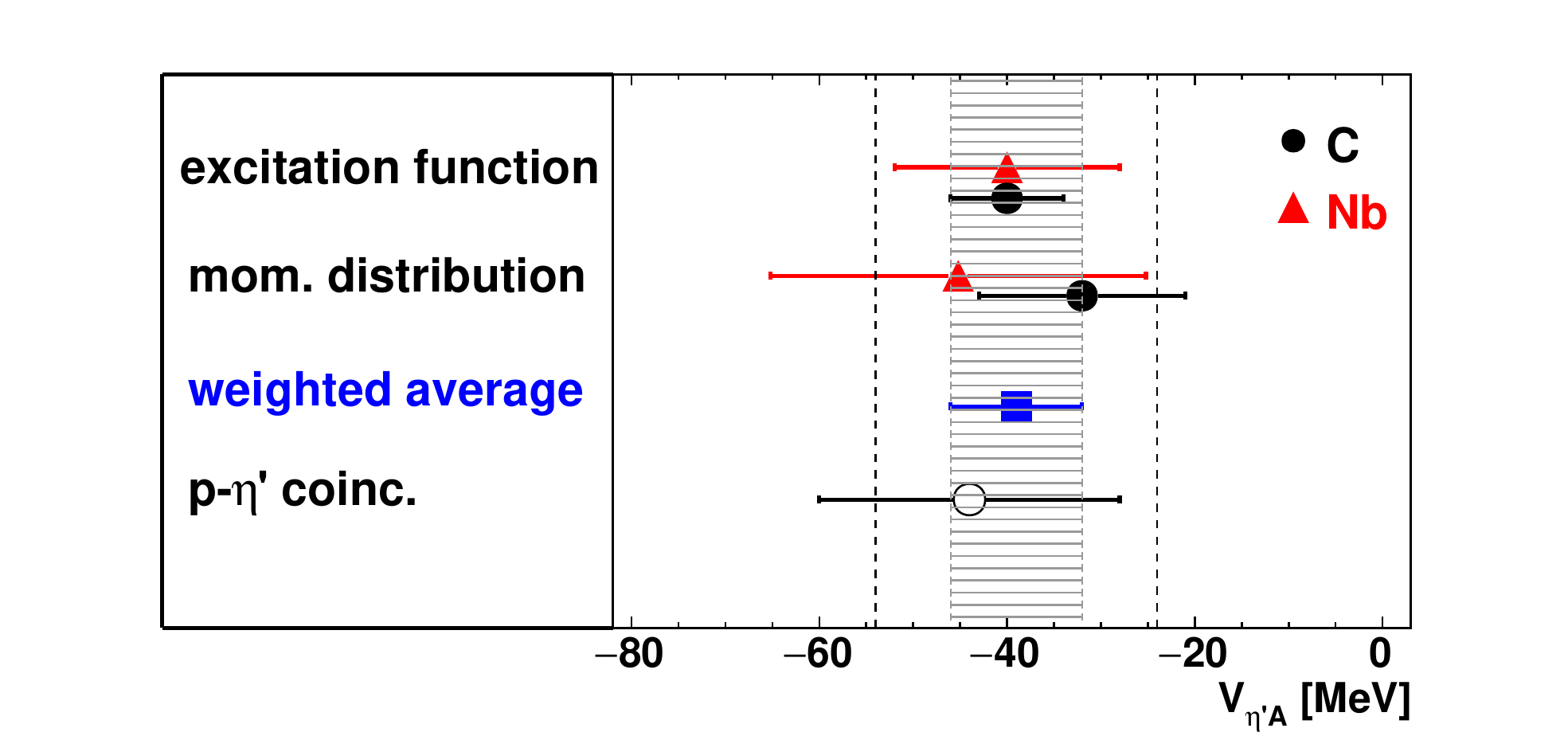}  \includegraphics[width=9.0cm,clip]{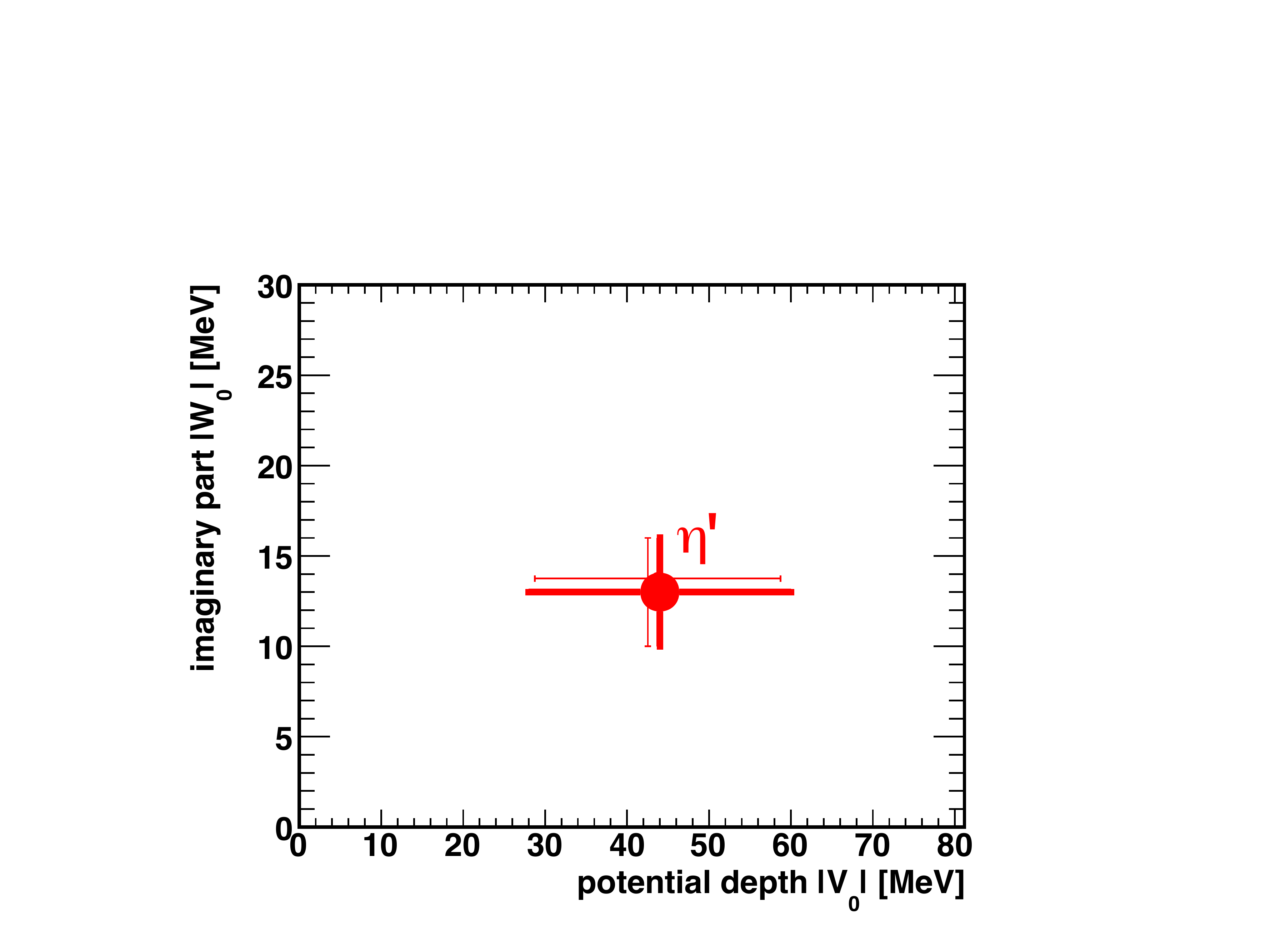}}
 \caption{Depth of the real part of the $\eta^\prime$-nucleus potential (left), determined for low $\eta^\prime$ momenta in this work (open black circle labelled p-$\eta^\prime$ coinc.), in comparison to previous determinations in inclusive experiments on carbon (full black circles) \cite{Nanova_C} and niobium (red triangles) \cite{Nanova_Nb}.  The weighted overall average is indicated by a blue square \cite{Metag_PPNP}. The shaded area indicates the statistical error. The hatched lines mark the range of systematic uncertainties. The real and imaginary part of the $\eta^\prime$-nucleus potential (right), determined for low $\eta^\prime$ momenta (this work and from \cite{Friedrich_TA}). Thick (thin) error bars correspond to statistical (systematic) errors, slightly shifted for better visibility.}
 \label{fig:compilation}
\end{center}
\end{figure*}

The measured cross sections are compared in Fig.~\ref{fig:Ekin_Exfunc_C} to calculations within the collision model \cite{Paryev_peta}. Using the measured differential cross sections for $\eta^\prime$ production off the proton and neutron bound in the deuteron \cite{Jaegle} as input, the cross section for $\eta^\prime$ photoproduction off C is calculated in an eikonal approximation, taking  the effect of the nuclear $\eta^\prime$ mean-field potential into account. The off-shell differential cross section for the production of $\eta^\prime$ mesons with reduced (or increased) in-medium mass off intranuclear protons in the elementary reactions $\gamma p \rightarrow \eta^\prime p$ is assumed to be given by the measured on-shell cross section, using the modified in-medium $\eta^\prime$ mass. The $\eta^\prime$ final-state absorption is taken into account by using a momentum independent, inelastic in-medium $\eta^\prime N$ cross section of $\sigma_{\eta^\prime N}$=13 mb, consistent with the recent result of transparency ratio measurements \cite{Friedrich_TA}. The contribution of $\eta^\prime$ production from two-nucleon short-range correlations is implemented by using the total nucleon spectral function in the parametrisation by \cite{Efremov}. As in \cite{Nanova_C,Nanova_Nb}, the momentum-dependent potential from \cite{Rudy}, seen by the protons emerging from the nucleus in coincidence with the $\eta^\prime$ mesons, is accounted for. The model calculation also considers coincident $\eta^\prime$ - proton production in a two-step process: the $\eta^\prime$ meson is produced off a first nucleon which then scatters off another nucleon of the carbon nucleus, kicking the final state proton into the MiniTAPS acceptance. It is found \cite{Paryev_peta} that such a secondary process is suppressed relative to primary $\eta^\prime$-proton events by more than an order of magnitude. A process involving a primary $\eta^\prime$ production off a proton with subsequent scattering of this proton off another target proton into MiniTAPS is suppressed also experimentally by the event selection, requiring one and only one charged hit in the whole detector system. The overall systematic uncertainties of the calculations are mainly given by the experimental input and the fits to the measured cross sections and are estimated to be of the order of 10-15$\%$.

These calculations are conceptually identical to the ones used for extracting the real part of the $\eta^\prime$-C potential in the inclusive reaction \cite{Nanova_C,Paryev}.  Here, however, the coincidence requirement of protons going into the polar range of $2^{\circ}-11^{\circ}$ with a minimum kinetic energy of 50 MeV is implemented, corresponding to the  proton detection threshold in MiniTAPS. Details of these calculations are described in \cite{Paryev_peta}. Furthermore it is required - as in the data  analysis - that the $\eta^\prime$ mesons go backwards in the center-of-mass system of the incident photon beam and a target nucleon at rest. The calculations have been performed for nine different scenarios assuming an $\eta^\prime$ real potential at normal nuclear matter density of $V$= +50, +25, 0, $-$25, $-$50, $-$75, $-$100, $-$125 and $-$150 MeV, respectively. The cross section curves calculated for the different scenarios are more clearly separated in case of the excitation functions which are thus more sensitive for the determination of the $\eta^\prime$ - nucleus potential. The calculations do not extend beyond E$_{\gamma}$ = 2.4 GeV since the elementary $\eta^\prime $ photoproduction cross sections off the free proton as well as off the proton and neutron bound in deuterium \cite{Crede,Jaegle} are only known up to this energy. The data and calculations are compared on an absolute scale, i.e., there is no rescaling of the calculations to the data.

Because of the limited statistics a simultaneous $\chi^{2}$-fit of the calculated cross section curves to the energy differential cross section (Fig.~\ref{fig:Ekin_Exfunc_C} (left)) and excitation function data (Fig.~\ref{fig:Ekin_Exfunc_C} (right)) has been performed, yielding a depth of the real part of the $\eta^\prime$ - carbon potential of  $-$(44 $\pm$ 16) MeV. By scaling the calculations up and down by 10$\%$, reflecting the systematic uncertainties of the calculations, we obtain a systematic error of $\pm$ 15 MeV. This value of the potential depth obtained for average $\eta^\prime $ momenta of $\approx 600 $ MeV/$c$ is consistent with earlier determinations of the real part of the $\eta^\prime$-nucleus potential in inclusive measurements \cite{Nanova_C,Nanova_Nb} as shown in Fig.~\ref{fig:compilation}. Thus, within the statistical and systematic uncertainties of this experiment there is no indication for a momentum dependence of the $\eta^\prime$ - nucleus potential. 

Comparing the real potential of $-$(44$\pm$16(stat)$\pm$15(syst)) MeV determined in this work to the imaginary potential of  $-$(13$\pm$3(stat)$\pm$3(syst)) MeV \cite{Friedrich_TA}, obtained by extrapolating the weak energy dependence of the imaginary part to the production threshold, we find that the modulus of the imaginary part - taking the experimental errors into account - is about a factor 2-4 smaller than the modulus of the real part also at relatively small  $\eta^\prime$ momenta, as shown in Fig.~\ref{fig:compilation}(right). This confirms that the $\eta^\prime$ meson is a suitable candidate for the observation of mesic states. Although the imaginary part of the potential appears to be quite small, allowing for the observation of relatively narrow states, it should be noted that the depth of the real potential is found to be much smaller than initially predicted theoretically \cite{Zaki} which may considerably reduce the strength of these states.

A recent search for $\eta^\prime \otimes$$^{11}$C bound states has not revealed any narrow mesic states, but the upper limit of about $-$100 MeV for the depth of the $\eta^\prime$ nucleus potential deduced from that experiment \cite{Tanaka} is consistent with the value reported in this work. For a potential depth of $\approx$~$-$40 MeV more sensitive measurements are required, e.g. by suppressing the background from multi-pion events by searching not only for the formation but in addition also for the characteristic decay of $\eta^\prime$- nucleus bound states.

\section{Conclusions}
The kinetic energy distribution and the excitation function for $\eta^\prime$ photoproduction off carbon and the free proton have been measured in coincidence with forward-going protons (2$^{\circ} \le \theta_{p} \le 11^{\circ}$). The data taken on the LH$_2$ target are in good agreement with model calculations and recent experimental results demonstrating the reliability of the analysis procedure and the calculations. A comparison of the carbon data with collision model calculations, performed under the experimental conditions, yields a real potential of $-$(44$\pm16$(stat)$\pm15$(syst)) MeV for the $\eta^\prime$ - carbon interaction for $\eta^\prime$ mesons with an average momentum of 600 MeV/$c$. This result is consistent with earlier determinations of the $\eta^\prime$ - nucleus potential \cite{Nanova_C,Nanova_Nb} in inclusive experiments for average $\eta^\prime$ momenta of 1.1 GeV/$c$. Within the experimental uncertainties  there is no indication for a momentum dependence of the $\eta^\prime$ nucleus interaction. Since the modulus of the imaginary potential near threshold is about a factor of 2-4 smaller than the modulus of the real part of the potential, the $\eta^\prime$ meson remains a good candidate for the search for meson-nucleus bound states although a first missing mass spectroscopy experiment \cite{Tanaka} has not identified narrow bound states. The potential depth of about $-$40 MeV, deduced in our photoproduction experiments, requires more sensitive measurements. A promising approach is to combine the missing mass spectrometry with detecting the decay of the $\eta^\prime$ mesic states in a semi-exclusive experiment. Promising decay channels appear to be two nucleon absorption or $\eta^\prime N \rightarrow \eta N$. Corresponding plans are pursued at the FRS@GSI/FAIR \cite{Itahashi_FRS} and by using photonuclear reactions \cite{Muramatsu,Metag}.

\section{Acknowledgements}
We thank the scientific and technical staff at ELSA and the collaborating
institutions for their important contribution to the success of the
experiment. Detailed discussions with U. Mosel are highly acknowledged. This work was supported financially by the {\it
 Deutsche Forschungsgemeinschaft} within SFB/TR16 and by the {\it Schweize\-ri\-scher Nationalfonds}. V. Crede acknowledges support from the U.S. Department of Energy.

\end{document}